\documentclass[twocolumn,prc,aps,showpacs,preprintnumbers,amsmath,amssymb]{revtex4}
\usepackage{graphicx}
\usepackage{appendix}

 \def\gsim{\mathrel{\rlap{\lower4pt\hbox{\hskip1pt$\sim$}}
 \raise1pt\hbox{$>$}}}

 \newcommand\la{\langle}
 \newcommand\ra{\rangle}
 \newcommand\beq{\begin{equation}}
 \newcommand\noi{\noindent}
 \newcommand\eeq{\end{equation}}
 \newcommand\beqn{\begin{eqnarray}}
 \newcommand\eeqn{\end{eqnarray}}
\def\mb{\,\mbox{mb}}
\def\fm{\,\mbox{fm}}
\def\GeV{\,\mbox{GeV}}

\def\lsim{\mathrel{\rlap{\lower4pt\hbox{\hskip1pt$\sim$}}
    \raise1pt\hbox{$<$}}}         
\def\gsim{\mathrel{\rlap{\lower4pt\hbox{\hskip1pt$\sim$}}
    \raise1pt\hbox{$>$}}}         

\def\mb{\,\mbox{mb}}
\def\fm{\,\mbox{fm}}
\def\GeV{\,\mbox{GeV}}

\def\s0{\sigma_0(s)}

\def\beq{\begin{equation}}
\def\eeq{\end{equation}}

\def\beqy{\begin{eqnarray}}
\def\eeqy{\end{eqnarray}}

\newcommand{\ber}{\begin{displaymath}}
\newcommand{\eer}{\end{displaymath}}
\newcommand{\bey}{\begin{eqnarray}}
\newcommand{\eey}{\end{eqnarray}}

\newcommand{\p}{\partial}

\def\beq{\begin{equation}}
\def\eeq{\end{equation}}

\def\beqy{\begin{eqnarray}}
\def\eeqy{\end{eqnarray}}

\begin{document}
\date{today}

\title{\bf Measuring the saturation scale in nuclei}
\author{B. Z. Kopeliovich}
\author{I. K. Potashnikova}
\author{Iv\'an Schmidt}
\affiliation{Departamento de F\'{\i}sica, Centro de Estudios
Subat\'omicos, Universidad T\'ecnica Federico Santa Mar\'{\i}a,
\\and\\
Centro Cient\'ifico-Tecnol\'ogico de Valpara\'iso,\\
Casilla 110-V, Valpara\'iso, Chile}

\begin{abstract}
\noi
The saturation momentum seeing in the nuclear infinite momentum frame is directly related to 
transverse momentum broadening of partons propagating through the medium  in the nuclear rest frame. Calculation of broadening within the color dipole approach including the effects of saturation in the nucleus, gives rise to an equation which describes well data on broadening in Drell-Yan reaction and heavy quarkonium production.

\end{abstract}

\date{\today}


\maketitle

\section{Introduction}

The theoretical description of interactions with nuclei frequently risks of breaking the unitarity bound.
Indeed in Born (impulse) approximation the cross section of any process on a nucleus is $A$ (or $A^2$ if the nucleus remains intact) times larger than that on a nucleon target. It might be still acceptable for hard reactions having a tiny cross section, very far from any unitarity constraints. However, amplitudes of soft processes are close to the unitarity bound, and multiplication by$A$ will easily violate unitarity.

This problem was first addressed and solved by Glauber \cite{glauber}: when an interaction is getting sufficiently strong, it starts screening itself. In particular, radiation of soft gluons, which has large cross section, also may be strongly shadowed. In this case one can interpret shadowing also in terms of the Landau-Pomeranchuk principle \cite{lp}: if the coherence time of radiation considerably exceeds the size of the target, the radiation process does not resolve between single and multiple interactions, only the total momentum transfer matters. In terms of the Fock state decomposition, a gluon in a given Fock state can be radiated only once.
Therefore, the spectrum radiated with small transverse momenta, $k_T^2\ll\la k_T^2\ra$, from multiple interactions must saturate when the phase space of radiated gluons is densely packed. Only at sufficiently large transverse momentum of gluons, $k_T\gsim\la k_T\ra$,
where the phase space becomes dilute, multiple interactions
start contributing to the multiplicity of gluons, increasing the range of $k_T$. Eventually, one arrives at
the Bethe-Heitler regime of radiation when each of multiple interactions equally contributes to the radiation spectrum.
The transverse gluon momentum characterizing the transition scale between the two regimes is called saturation momentum and is defined below. 

The same phenomenon of saturation looks different in the infinite momentum frame of the nucleus.
If the bound nucleons do not overlap in the nuclear rest frame, they should stay separated after a boost to the infinite momentum frame either, since Lorentz contractions affect the nucleon and the inter-nucleon spacing in the same way. However, as a result of Lorentz boost the nucleons acquire long living vacuum fluctuations of the Weitz\"acker-Williams type, which are identified as partons.
Partons carrying a small fraction of the nucleon momentum, $x\ll1$, are Lorentz contracted much less than the most energetic part of the proton, therefore the parton clouds originated from different nucleons overlap in the longitudinal direction at small $x$
and start interacting \cite{kancheli}. This overlap also leads to a significant increase of the parton density in the impact parameter plane, especially at small momenta of gluons where the density is rather large eve in a single nucleon. However, interferences reduce the gluon density \cite{mv} similar to how the LP effect suppresses gluon radiation with small $k_T$. As a result, the mean transverse momentum of small-$x$ gluons in a nucleus is pushed up to higher value called saturation momentum. This effect is known nowadays under the name color glass condensate (CGC) \cite{cgc1,cgc2}.
This phenomenon can be also interpreted in terms of the parton model as parton fusion leading to saturation of the parton density \cite{glr}.  

We evaluate the saturation momentum within the dipole approach, and
derive an equation which involves the effect of saturation. This makes the treatment of gluon shadowing self-consistent and leads to a considerable reduction of the saturation scale. We found that the saturation scale becomes independent of $A$ for very (unrealistically)
heavy nuclei.

We relate the saturation scale to the experimentally observed broadening of transverse momentum in different processes, and test different models. The dipole approach agrees well with data, while
other models relating the saturation momentum to the gluon density at the saturation scale, considerably overestimate data.

\section{How to measure the saturation momentum}\label{broadening}

The rise of total cross sections with energy, discovered back in 1973, was the first manifestation of
an increasing population of partons towards smaller $x$. As the parton density increases,
the inverse process of parton fusion becomes important, and eventually the parton density is expected \cite{glr} to saturate. The related phenomenon, called nowadays color glass condensate, is an increase of the mean transverse momentum of the partons up to a characteristic value $Q_A$, called saturation momentum. 

In the nuclear rest frame the same phenomenon looks like Glauber shadowing and color filtering for a dipole (quark-antiquark, or glue-glue) of transverse separation $r_T$ and energy $E$ propagating through a nuclear matter \cite{al}.
The partial elastic dipole-nucleus amplitude at impact parameter $b$ reads \cite{zkl},
\beq
f_{\bar qq}^A(b)=1-e^{-{1\over2}\sigma_{\bar qq}^N(r_T,E)\,T_A(b)},
\label{100}
\eeq
where $T_A(b)=\int_{-\infty}^{\infty} dz\,\rho_A(b,z)$ is the nuclear thickness function, integral of nuclear density along the trajectory of the projectile at impact parameter $b$. 
We assume here that the Lorentz delated length (coherence length) of the dipole size fluctuations is much longer than the nucleus.
The $\bar qq$ dipole cross section on a nucleon should vanish at small transverse separation
$\sigma_{\bar qq}(r_T\to0,E)\propto r_T^2$.  The energy dependence is discussed below.
For large $ T_A(b)$ only small-$r_T$ part of the cross section contributes, so one can use the $r_T^2$-approximation, 
\beq
\sigma_{\bar qq}^N(r_T,E)\approx C_q(E,r_T)\,r_T^2,
\label{120}
\eeq
where $C_q(E,r_T)$ is logarithmically divergent at small $r_T$ \cite{zkl}. In this limit the factor $C_q$
can be related to the gluon distribution \cite{fs},
\beq
C_q(E,r_T)=\frac{\pi^2}{3}\,\alpha_s(1/r_T^2)\,xG(x,1/r_T^2),
\label{130}
\eeq
where $1/x=2m_NE\,r_T^2 $.

One can introduce the quark saturation momentum $Q_{qA}$  presenting Eqs.~(\ref{100})-(\ref{120}) as,
\beq
f_{\bar qq}^A(b)=1-\exp\left[-\frac{r_T^2\,Q_{qA}^2(b,E)}{4}\right],
\label{140}
\eeq
where 
\beq
Q_{qA}^2(b,E)=2C_q(E,r_T=1/Q_{qA})T_A(b).
\label{160}
\eeq
Here we fixed the dipole separation at the typical value $r_t\sim1/Q_{qA}$ relying on the weak, logarithmic, $r_T$-dependence of $C(E,r_T)$.

The same value as in Eq.~(\ref{160}) controls broadening of transverse momentum of a single parton propagating through a nucleus \cite{dhk,jkt},
\beq
\Delta p_T^2=2T_A(b)
\left.\frac{d\sigma_{\bar qq}^N(r_T)}{dr_T^2}\right|_{r_T=0}=
2T_A(b)C_q(E,r_T=0).
\label{180}
\eeq
Thus, we arrive at a divergent result, since $C(E,r_T)\propto \ln(1/r_T)$ at $r_T\to0$ \cite{zkl}.
This is not a surprise, as the mean transverse momentum squared $\la p_T^2\ra$ is ultra-violet divergent.
Moreover, this divergency is not cancelled in broadening $\Delta p_T^2=\la p_T^2\ra_A-
\la p_T^2\ra_N$ \cite{mikkel-raufeisen,kovner-wied}.
To settle the problem one should fix $r_T$ at a characteristic value for the process under consideration,
i.e. at $r_T^2\sim 1/\Delta p_T^2$.

Thus, broadening and the saturation momentum are equal,
\beq
Q_{qA}^2(b,E)=\Delta p_T^2(b,E),
\label{190}
\eeq
so one has a direct access to the saturation scale by measuring broadening.

Notice that the approximation we used in Eqs.~(\ref{160}) and (\ref{180}), neglecting the weak $r_T$ dependence of $C_q(E,r_T)$, is well justified by data. Indeed, the parametrization
of the dipole-proton cross section \cite{gbw}, which has finite $C_q(E,r_T=0)$, describes reasonably well the small-$x$ DIS data up to very high virtualities,  $Q^2\sim100\GeV^2$, which is much larger than the saturation scale. Of course, introduction of an additional $r_T$ dependence via gluon density like in (\ref{130}) improves agreement with data. However, it leaves broadening divergent.
In what follows we will neglect the dependence of $C(E)$ on $r_T$, employing the parametrizations presented in Refs.~\cite{gbw,kst2}.

\subsection{Quark broadening}

Broadening is predominantly a soft interaction process, and the transverse momentum increases as a result of many soft rescatterings. Even if $Q_{qA}^2$ is large, it is not correct to use an unintegrated gluon density of the nucleus at this scale.
Broadening is a result of multiple soft gluon exchanges, rather than a single scattering, with a distribution given by the nuclear unintegrated gluon density. 
Although a parton-nucleon differential cross section is infra-red divergent, the result of broadening is finite(see the Moli\`ere theory of multiple interactions in \cite{jkt}). Therefore, to evaluate the factor $C_q(E)=\partial\sigma_{\bar qq}(r_T,E)/\partial r_T^2\bigr|_{r_T=0}$ in (\ref{120}), one has to know the dipole cross section fitted to soft processes, rather than to DIS. Correspondingly, this cross section should depend on energy, rather than on Bjorken $x$. Such a cross section, parametrized in the saturated form,
\beq
\sigma^N_{\bar qq}(r_T,E)=\sigma_0(E)\left[1-e^{-r_T^2\,Q_{qN}^2(E)/4}\right],
\label{200}
\eeq
and fitted to data for the $\pi N$ total cross section, photoproduction of vector mesons and DIS data for not high $Q^2<10\GeV^2$, results in the following parameters,
\beqn
Q_{qN}(E) &=& 0.19\GeV\times\left(\frac{E}{1GeV}\right)^{0.14}
\label{220}\\
\sigma_0(E) &=&
\sigma^{\pi p}_{tot}(E)\left[1+\frac{3}{2\, Q_{qN}^2(E)\, \la r_{ch}^2\ra_\pi}\right]
\label{240}
\eeqn
The Pomeron part of the $\pi p$ total cross section is parametrized as $\sigma^{\pi p}_{tot}(E)=
14.35\mb\times(E/1\GeV)^{0.08}$.
Thus, the factor $C_q(E)$ in (\ref{120}) has the form,
\beqn
C_q(E)&\equiv&\frac{\partial\sigma_{\bar qq}^N(r_T,E)}{\partial r_T^2}\biggr|_{r_T=0} 
\nonumber\\ &=&
{1\over4}\,\sigma^{\pi p}_{tot}(E)\,\left[Q_{qN}^2(E)
+\frac{3}{2\, \la r_{ch}^2\ra_\pi}\right]
\label{260}
\eeqn
Notice that at low energies the second term in square brackets dominate and the broadening slowly rises with energy, as $E^{0.08}$. Then,
with $\la r_{ch}^2\ra_\pi=0.44\fm^2$, the two terms in square brackets become equal at the energy of about $100\GeV$. At higher energy the first term takes over and at high energies broadening steeply rises, as $E^{0.36}$.

\begin{figure}[htb]
 \includegraphics[width=8cm]{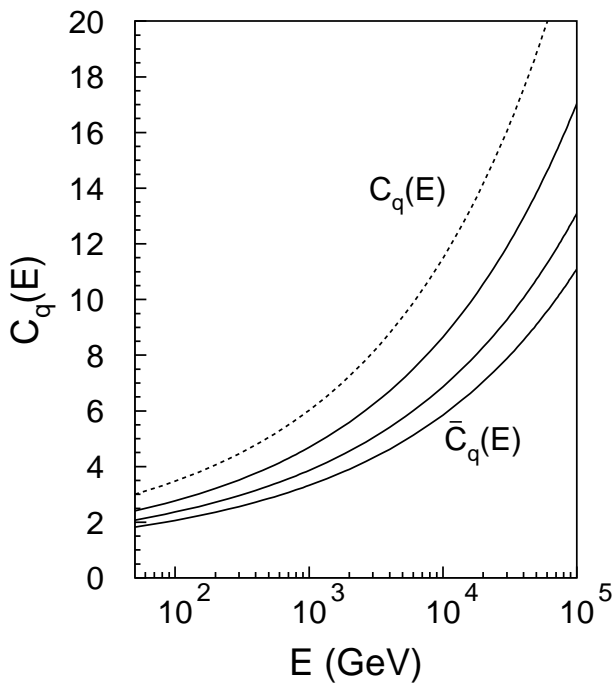}
 \caption{\label{e-dep} Dashed curve shows $C_q(E)$ calculated with Eq.~(\ref{260}) as function of quark energy. Solid curves show the modified broadening factor $\widetilde C_q(E)=R_g(E)\,C_q(E)$, damped down by gluon shadowing, which depends on nuclear thickness $T_A$ propagated by the quark. The curves from bottom to top correspond to $T_A=1.5,\ 1.0,\ 0.5\fm^{-2}$.}
 \end{figure}

The energy dependence of $C_q(E)$, calculated with Eq.~(\ref{260}), is depicted in Fig.~\ref{e-dep}.
 
 Notice that as far as the shape of the dipole cross section Eq.~(\ref{200}) is known, one can calculate not only broadening, but the whole transverse momentum distribution of quarks propagating through a nucleus at impact parameter $b$ \cite{jkt},
   \beqn
  && \frac{dN_q(b)}{d^2p_T} =
   \int d^2r_1d^2r_2\,
   e^{i\vec p_T\cdot(\vec r_1-\vec r_2)}\,
   \Omega^q_{in}(\vec r_1,\vec r_2)
   \nonumber\\ &\times&
   \exp\left[-\sigma_{\bar qq}(\vec r_1-\vec r_2,E)\,
   T_A\left(\vec b+\frac{\vec r_1+\vec r_2}{2}\right)\right]
   \label{265}
   \eeqn
   Here $\Omega^q_{in}(\vec r_1,\vec r_2)$ is the density matrix describing the spacial distribution of the projectile quark in the incoming hadron.

 \subsection{Broadening of  photons and dileptons}\label{photons}
 
Apparently, a single photon propagating through a medium does not experience any broadening.  Nevertheless, calculations \cite{kst1,jkt} and data show that direct photons and dileptons are subject to the Cronin effect, which is a manifestation of broadening.  According to \cite{kst1}
the cross section of photon radiation with fractional light-cone momentum $\alpha$ by a quark propagating through a nucleus at impact parameter $b$, integrated over the transverse momentum of the recoil quark, reads,
\beqn
&&\frac{d\sigma_A(q\to\gamma q)}{d^2b\,d\ln\alpha\,d^2p_T}=
\frac{1}{(2\pi)^2}
\int d^2r_1 d^2r_2\,e^{i\vec p_T\cdot(\vec r_1-\vec r_2)}
\nonumber \\ &\times&
\Psi_{q\gamma}^\dagger(\alpha,\vec r_1)\,
\Sigma^A_\gamma(\vec r_1,\vec r_2,\alpha,b)\,
\Psi_{q\gamma}(\alpha,\vec r_2),
\label{272}
\eeqn
where
$\Psi_{q\gamma}(\alpha,\vec r)$ is the light-cone distribution function for the $|q\gamma\ra$ Fock component of the quark, defined in \cite{hir,kst1}; 
\beqn
&&
\Sigma^A_\gamma(\vec r_1,\vec r_2,\alpha,b)=1 +
e^{-{1\over2}\sigma_{\bar qq}[\alpha(\vec r_1-\vec r_2)]\,T_A(b)}
\nonumber\\ &-&
e^{-{1\over2}\sigma_{\bar qq}(\vec r_1)\,T_A(b)} -
e^{-{1\over2}\sigma_{\bar qq}(\vec r_2)\,T_A(b)}.
\label{274}
\eeqn
Notice that Eq.~(\ref{272}) is valid only at high energies where the coherence length of photons radiation considerably exceeds the nuclear size, $l_c\gg R_A$, where
\beq
l_c=\frac{2\alpha(1-\alpha)E}{k_T^2+(1-\alpha)M^2},
\label{275}
\eeq
and $M$ is the photon mass (dilepton invariant mass in Drell-Yan reaction).
In this limit the broadening calculated with (\ref{272}) reads,
 \beq
 \left(\Delta p_T^2\right)_{\gamma}^{l_c\gg R_A}=2\alpha^2 C_q(E)\,T_A(b),
 \label{276}
 \eeq
 In this case the observed broadening is not the result of propagation of a single photon through the nucleus, but  is due to the nuclear modification of the whole radiation process, which happens due to interaction of the source quark.

In another limiting regime of coherence length shorter than the mean internucleon spacing in the nucleus, $l_c\lsim 1\fm$, the photon is radiated instantaneously inside the nucleus, and broadening occurs due initial state interactions and broadening of the source quark. Therefore, one can rely on broadening given by Eq.~(\ref{180}), remembering that the path length available for initial state interactions is twice shorter than in (\ref{180}), and the radiated photon carries only fraction $\alpha$ of the quark transverse momentum. Thus, we arrive to a new expression for broadening,
\beq
\left(\Delta p_T^2\right)_{\gamma}^{l_c\lsim1\fm}=\alpha^2 C_q(E)\,T_A(b),
 \label{277}
 \eeq
 which is similar to (\ref{276}), but is twice smaller.

Gluon radiation should be treated in a similar way \cite{kst1,kst2}.

\subsection{Broadening of gluons}

Propagation of a gluon through a medium looks similar to that for a quark, except that the effect should be stronger, since the cross section of a glue-glue dipole at small separations is enhanced by the Casimir factor compared with a $\bar qq$ dipole, so gluon broadening has the form,
 \beq
\left(\Delta p_T^2\right)_g=2 C_g(E)\,T_A(b)={9\over2}\,C_q(E)\,T_A(b).
 \label{270}
 \eeq
 
 However, as we have just seen on the example of photon radiation, one should consider broadening for the whole radiation process, rather than for propagation of a single photon.  
 
 Broadening for a radiated gluon depends on the coherence length, Eq.~(\ref{275}), where $M=m_g$ is the effective gluon mass. This mass (or the mean gluon transverse momentum) may be introduced in order to take into account nonperturbative QCD effects in the quark-gluon light cone wave function, and is fixed by data at $m+g\approx 0.65\GeV$ \cite{kst2,spots}.
In the limit of long coherence length, $l_c\gg R_A$, the cross section of gluon radiated with fractional light-cone momentum $\alpha$ by a quark propagating through a nucleus at impact parameter $b$, integrated over the transverse momentum of the recoil quark, reads \cite{kst1},
\beqn
&&\frac{d\sigma_A(q\to gq)}{d^2b\,d\ln\alpha\,d^2p_T}=
\frac{1}{2\pi^2}
\int d^2r_1 d^2r_2\,e^{i\vec p_T\cdot(\vec r_1-\vec r_2)}
\nonumber \\ &\times&
\Psi_{qg}^\dagger(\alpha,\vec r_1)\,
\Sigma^A_g(\vec r_1,\vec r_2,\alpha,b)\,
\Psi_{qg}(\alpha,\vec r_2),
\label{272a}
\eeqn
where
$\Psi_{qg}(\alpha,\vec r)$ is the light-cone distribution function for the $|qg\ra$ Fock component of the quark, defined in \cite{kst1,kst2}; 
\beqn
&&
\Sigma^A_g(\vec r_1,\vec r_2,\alpha,b)=
e^{-{1\over2}\sigma_{\bar qq}[\alpha(\vec r_1-\vec r_2)]\,T_A(b)}
\nonumber\\ &+&
e^{-{1\over2}\sigma_{gg}(\vec r_1-\vec r_2)\,T_A(b)} -
e^{-{1\over2}\sigma_{g\bar qq}(\vec r_1,\vec r_1-\alpha\vec r_2)\,T_A(b)}
\nonumber\\ &-&
e^{-{1\over2}\sigma_{g\bar qq}(\vec r_2,\vec r_2-\alpha\vec r_1)\,T_A(b)};
\label{274a}
\eeqn
 and $\sigma_{g\bar qq}(\vec\rho_1,\vec\rho_2)={9\over8}[\sigma_{\bar qq}(\rho_1)+\sigma_{\bar qq}(\rho_2)]-{1\over8}\sigma_{\bar qq}(\vec\rho_1-\vec\rho_2)$ is the cross section of a 3-body dipole consisted of a gluon, quark and antiquark, with transverse separations $\vec\rho_1$ and $\vec\rho_2$
 between the gluon and $q$, or $\bar q$ respectively.
 
 Then one can calculate the mean values of $p_T^2$ for a proton and nuclear targets and subtracting 
them get broadening. The last two terms in (\ref{274a}) lead to an exponentially falling $T_A$-dependence, so we neglect them. The rest gives,
 \beq
 \left(\Delta p_T^2\right)_{q\to gq}^{l_c\gg R_A} = 2
\left(\frac{9}{4}+\alpha^2\right)C_q(E)\, T_A(b),
\label{278a}
\eeq
which contains an additional term of the order $\alpha^2$ compared to Eq.~(\ref{270}).

In the limit of short coherence length, $l_c\lsim 1\fm$, broadening for gluon radiation has two sources: (i) broadening of the projectile quark due to initial state interactions; (ii) broadening of the radiated gluon due to final state interaction. Correspondingly, the amounts of broadening coming from these two sources read,
\beqn
 \left(\Delta p_T^2\right)_{q\to gq}^{l_c\lsim 1\fm} &=& 
 2\alpha^2C_q(E)\, T_z(b)\ \  {\rm (initial\ state)};
 \label{278b}\\
 \left(\Delta p_T^2\right)_{q\to gq}^{l_c\lsim 1\fm} &=& 
 {9\over2}C_q(E)[T_A(b)-T_z(b)]\ \  {\rm (final\ state)}.
\nonumber 
\eeqn
Here $T_z(b)=\int_{-\infty}^z dz'\,\rho_A(b,z)$ is the nuclear thickness passed by the projectile quark before the gluon radiation occurs at the point $z$ with a short $l_c$. Summing up the two contributions and averaging over $z$ we eventually arrive at,
 \beq
 \left(\Delta p_T^2\right)_{q\to gq}^{l_c\lsim 1\fm} = 
\left(\frac{9}{4}+\alpha^2\right)C_q(E)\, T_A(b),
\label{278c}
\eeq
Amazingly, the amounts of broadening of radiated gluons in the two regimes of radiation, long and short $l_c$, are different by factor of two, similar to what was found for broadening of photons.

Radiation of gluons at small $x\ll1$ is dominated by the gluon splitting process $g\to gg$, rather than by direct radiation by the valence quarks, $q\to gq$, which is suppressed by powers of $\ln(1/x)$.
Broadening of the radiated gluons in the two limiting regimes of coherent and incoherent radiation is given by,
\beqn
\left(\Delta p_T^2\right)_{g\to gg}^{l_c\gg R_A} &=& {9\over2}
\left(1+\alpha^2\right)C_q(E)\, T_A(b);
\nonumber\\
\left(\Delta p_T^2\right)_{g\to gg}^{l_c\lsim 1\fm} &=& 
{1\over2}\left(\Delta p_T^2\right)_{g\to gg}^{l_c\gg R_A}
\label{278d}
\eeqn
The mean fractional momentum $\alpha$ of the radiated gluon calculated with the DGLAP splitting function is rather small,
\beq
\la\alpha^2\ra=\frac{1}{8\ln(1/x)},
\label{278e}
\eeq
therefore the difference between gluon broadening in the radiation processes, $q\to gq$, or $g\to gg$,
and broadening of a single gluon propagating through the nucleus is quite small, if the radiation occurs in the long coherence length regime, $l_c\gg R_A$. Otherwise the difference ma considerably depend on the fractional momentum of the radiated gluon.

 \section{Saturation in a saturated environment}\label{gluon-shad}
 
 The energy dependence of the dipole cross section originates from gluon radiation.
 The eikonal  expression Eq.~(\ref{100}) assumes that radiation occurs in the Bethe-Heitler regime, i.e. in every one of multiple collisions a full spectrum  of gluons is radiated. This may be true at low energies when the coherence time of radiation is shorter than the mean free path of the parton in the medium. However, at high energies gluon radiation from multiple interactions is subject to interferences,  which suppress the radiation rate. This coherence phenomenon is known as Landau-Pomeranchuk-Migdal (LPM) effect \cite{lp,m}, or can be interpreted in terms of gluon shadowing.
 
 Using Eq.~(\ref{160}) and the definition of the factor $C_g(E)$ in Eqs.~(\ref{270}) and (\ref{120}), we can relate the gluon saturation scale in a nucleus to the unintegrated gluon density,
 \beq
 Q_{gA}^2(E,b)=
  \frac{3\pi}{2}\,T_A(b)
 \int d^2k\,\frac{\alpha_s(k^2)}{k^2}\,
 {\cal F}_N(x,k^2),
 \label{279}
 \eeq
where ${\cal F}_N(x,k^2)$ is the unintegrated gluon density in a nucleon;
$x=k^2/2m_NE$. This equation contains nothing new so far, and it is equivalent to Eq.~(\ref{270}).
To incorporate the effects of coherence in gluon radiation into our calculations we rely on Eq.~(52) of Ref.~\cite{jkt}, and for the gluon saturation momentum modified by gluon shadowing we get,
 \beqn
 \widetilde Q_{gA}^2(E,b)&=&\frac{3\pi}{2}\,T_A(b)
 \int d^2k\,\frac{\alpha_s(k^2)}{k^2}\,
 {\cal F}_N(x,k^2)
 \nonumber\\ &\times&
 S_A(\widetilde Q_{gA}^2,k^2,b),
 \label{279a}
 \eeqn
The nuclear modification factor $S_A(\widetilde Q_{gA}^2,k^2,b)$ takes care of the LPM suppression, which is controlled by the same gluon saturation momentum $\widetilde Q_{gA}^2$ as in the left-hand-side of Eq.~(\ref{279a}). Thus, we arrived at the equation for the saturation scale modified by coherence effects.

\subsection{Parametrizing \boldmath$S_A$}\label{model-1}

The physics of suppression is rather intuitive. If the radiation length $l_c=2x(1-x)E/k^2$ of a gluon with certain $x$ and $\vec k$ considerably exceeds the nuclear size, multiple interactions cannot generate an identical gluon to be radiated within the same phase space cell. This is the very sense of saturation: the phase space for gluons with given $x$ is packed up to transverse momentum $k^2\lsim
\widetilde Q_{gA}^2$, and only above this bound new gluons can be generated by multiple interactions.
Correspondingly, we propose the following simple model for $S(x,k^2)$, which incorporates  these features,
\beq
S_A(E,k^2,b)=1-\frac{\sigma_{eff}T_A(b)}{1+\sigma_{eff}T_A(b)}\,
e^{-k^2/\widetilde Q_{gA}^2(E,b)}
\label{279b}
\eeq
where the effective radiation cross section $\sigma_{eff}$ specified later, controls the number of collisions contributing to gluon radiation.

Expression (\ref{279b}) interpolates between the two limiting regimes: (i) at $k^2\gg \widetilde Q_{gA}(E,b)^2$
the density of gluons is very low, their phase space is dilute and the radiated gluons do not interfere with each other. In this case $S(E,k^2,b)=1$; (ii) at $k^2\ll \widetilde Q_{gA}(E,b)^2$
the gluon density saturates and there is no room for radiation of extra gluons. Therefore,
the gluon radiation spectrum in this regime should be the same as in a single interaction.
We remind that equations (\ref{160}), (\ref{260}) were calculated in the Bethe-Heitler (BH) regime, when
the number of radiated gluons on a nucleus is proportional to the effective number of collisions,
\beq
\left.\frac{dn^g_A(b)}{d^2k}\right|_{BH}
=\left[1+\sigma_{eff}T_A(b)\right]
\frac{dn^g_N(b)}{d^2k},
\label{279c}
\eeq
where one is added, since at least one interaction must occur. Thus, in the regime of full coherence and saturation of gluons one should reduce the Bethe-Heitler spectrum of gluons by a factor $[1+\sigma_{eff}T_A]$. Eq.~(\ref{279b}) satisfies this limiting behavior. 

Employing the model Eq.~(\ref{200}) one gets for the unintegrated gluon density \cite{gbw},
\beq
{\cal F}_N(x,k^2)=\frac{3\sigma_0(E)\,k^4}{4\pi^2\alpha_s(k^2)Q_{qN}^2(E)}\,
e^{-k^2/Q_{qN}^2(E)},
\label{279d}
\eeq
and from Eq.~(\ref{279a})
\beq
\frac{\widetilde Q_{gA}^2(E,b)}{Q_{gA}^2(E,b)} = 
1\,-\,\frac{\sigma_{eff}T_A(b)}
{\left[1+\sigma_{eff}T_A(b)\right]\left[1+
Q_{qN}^2/\widetilde Q_{gA}^2\right]^2}
\label{279ae}
\eeq

Then, we arrive at an equation for the shadowing modified saturation scale. For the ratio of the modified to unmodified saturation momenta squared, 
\beq
R_g(E,b)\equiv\frac{\widetilde Q_{gA}^2(E,b)}{Q_{gA}^2(E,b)}=
\frac{\widetilde Q_{qA}^2(E,b)}{Q_{qA}^2(E,b)},
\label{279j}
\eeq
the equation can be represented in the form,
\beq
R_g=1-
\frac{R_g^2\,n_0^2\,n_{eff}}{(1+R_g\,n_0)^2(1+n_{eff})}
\label{279e}
\eeq
where 
\beqn
n_0(E,b)&=&{9\over8}\,\sigma_0(E)\,T_A(b);
\nonumber\\
n_{eff}(E,b)&=&\sigma_{eff}(E)\,T_A(b).
\label{279ee}
\eeqn
Here $\sigma_0(E)$ is given by Eq.~(\ref{240}) and $\sigma_{eff}(E)$ is specified below.

 Notice that in the limit of very large $T_A$, when $\sigma_{eff}T_A(b)\gg1$ and $R_g(E,b)\sigma_0(E)T_A(b)\gg1$ the shadowing suppression decreases as $R_g(E,b)\propto 1/T_A(b)$.
Correspondingly, the modified saturation scale $\widetilde Q_{qA}^2(E,b)$ becomes independent off $T_A$ and of $b$.

The effective cross section of gluon radiation $\sigma_{eff}$ deserves special attention.
There are many evidences in data indicating that this cross section is rather small \cite{spots}.
Probably the most appealing and direct experimental fact is the smallness of the diffractive gluon radiation. The observed large invariant mass behavior of the diffractive cross section $pp\to pX$, $d\sigma_{dd}/dM^2\propto 1/M^2$, is an explicit manifestation of radiation of a vector particle, i.e. a gluon. This cross section is an order of magnitude smaller than any simple expectation.
In terms of Regge theory this is known as the longstanding problem of smallness of the triple-Pomeron coupling. 

The only known reasonable solution for this puzzle is to assume that nonperturbative effects squeeze the glue-glue light-cone wave function down to a small mean separation $r_0\approx 0.3\fm$ \cite{kst2,spots}. Correspondingly the cross section for such a dipole, which controls gluon radiation reads,
\beq
\sigma_{eff}=C_g(E)\,r_0^2,
\label{279f}
\eeq
where $C_g(E)$ was introduced in (\ref{270}).
This cross section is rather small, a few mb. Correspondingly the shadowing modification effect is rather small, much smaller than is permitted by the unitarity bound \cite{klps-bound}.
This smallness is a result of rare overlap of small gluon clouds in impact parameter plane \cite{kst2,spots}.

\begin{figure}[htb]
 \includegraphics[width=8cm]{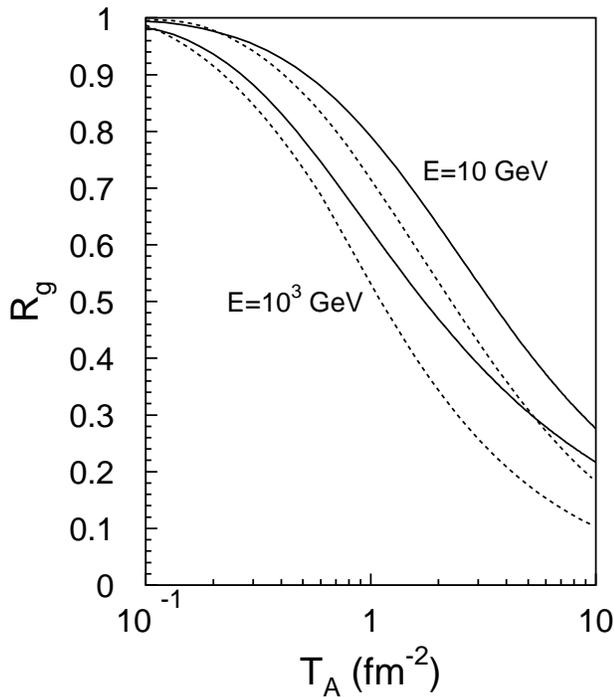}
 \caption{\label{Rg} Gluon shadowing ratio $R_g(E,b)$ as function of nuclear thickness $T_A$.
 Solid and dashed curves correspond to different parametrizations of the LPM suppression factor,
 Eqs.~(\ref{279b}) and (\ref{400}) respectively. Two upper and two bottom curves correspond to energies $E=10$ and $10^3\GeV$ respectively.}
 \end{figure}

\begin{figure}[htb]
 \includegraphics[width=8cm]{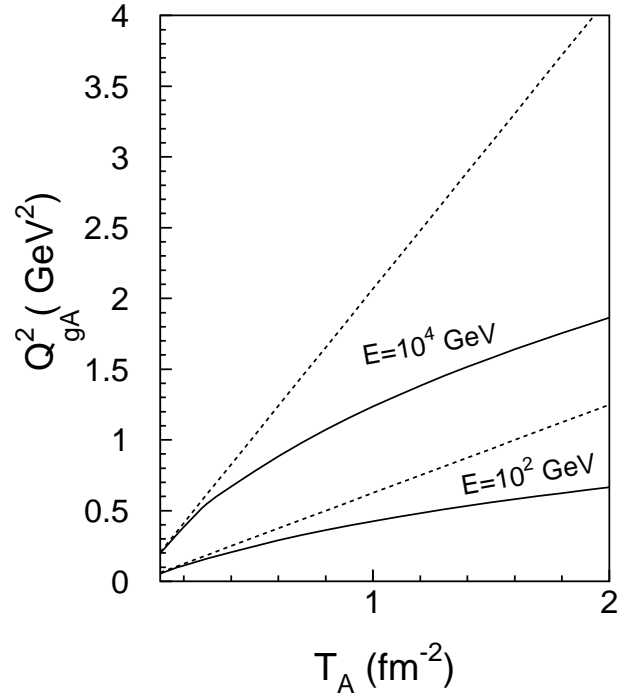}
 \caption{\label{sat-mom} The saturation momentum squared as function of nuclear thickness. 
 Dashed curves correspond to Bethe-Heitler regime of no shadowing }
 \end{figure}

The solution of Eq.~(\ref{279e}) for $R_g(E,b)$  is plotted in Fig.~\ref{Rg} as function of nuclear thickness, at different energies.
To demonstrate the effect of gluon shadowing we plot in Fig.~\ref{sat-mom}
both the unshadowed, $Q_{qA}(E,b)$, and shadowed, $\widetilde Q_{qA}(E,b)$, saturation scales as function of $T_A$. 
We see that the shadow corrected saturation momentum has a tendency to level off.
This is not a surprise, since the gluon shadowing corrections rise as $T_A^2$ \cite{mine,diffraction}.
Eventually, the saturation momentum becomes independent of $T_A$, as was expected in \cite{levin}, but we found that this happens only for unrealistically heavy nuclei, $T_A\sim 10\fm^{-2}$, while even heaviest nuclei have $T_A\lsim 2\fm^{-2}$.

The shadowing modified broadening factor  $\widetilde C_q(E,b)=R_g(E,b)\,C_q(E)$ is also plotted in Fig.~\ref{e-dep} for several values of $T_A$. While the factor $C_q(E)$ is independent of $T_A$ by the definition Eq.~(\ref{160}), the gluon shadowing corrections rise with $T_A$.

Concluding this section, a word of caution is in order. The above consideration of gluon shadowing is valid only if the coherence length of gluon radiation with energy $E$ and transverse momenta up to the saturation scale considerably exceeds the nuclear size,
\beq 
l_c=\frac{2E}{Q_{qA}^2}\gg R_A.
\label{279g}
\eeq

\subsection{Alternative shape of the LP suppression}\label{model-2}

In order to get a hint for theoretical uncertainties related to the model dependence of our calculation of gluon shadowing we test here another model for the LP suppression factor $S_A(\widetilde Q_{gA}^2,k^2,b)$ in Eq.~(\ref{279a}). As an alternative to the model given by Eq.~(\ref{279b}), one can consider another trial function,
\beq
S_A(\widetilde Q_{gA}^2,k^2,b) =
1 - \frac{\sigma_{eff}T_A(b)}{1+\sigma_{eff}T_A(b)}\,
\Theta(\widetilde Q_{gA}^2-k^2).
\label{400}
\eeq
This function has the same limits at small and large $k^2$ as the one Eq.~(\ref{279b}), but with a sharp transition at $k^2=\widetilde Q_{gA}^2$.

With such an LP suppression factor and the same unintegrated gluon density Eq.~(\ref{279d}) we arrive at a new equation for gluon shadowing, alternative to Eq.~(\ref{279e}),
\beq
R_g=\frac{1}{1+n_{eff}}\,\left[1+
n_{eff}(1+R_gn_0)\,
e^{-R_gn_0}\right]
\label{420}
\eeq
The results for $R_g$ as function of nuclear thickness $T_A$ at different energies is plotted in Fig.~\ref{Rg} by dashed curves.
The strength of suppression is rather similar to what is presented by solid curves corresponding to the parametrization Eq.~(\ref{279b}), demonstrating a weak dependence on the way of interpolating  between saturated and Bethe-Heitler regimes.

\section{Saturation scale from data}
 
  Broadening of partons in nuclear matter has been studied in several experiments with different processes at different energies \cite{mmp,e866,e772,e866-pt,e789,e866-psi,jlab,hermes}. Here we overview the results of these measurements.
 
  \subsection{Drell-Yan reaction}
 
\begin{figure}[htb]
 \includegraphics[width=7.5cm]{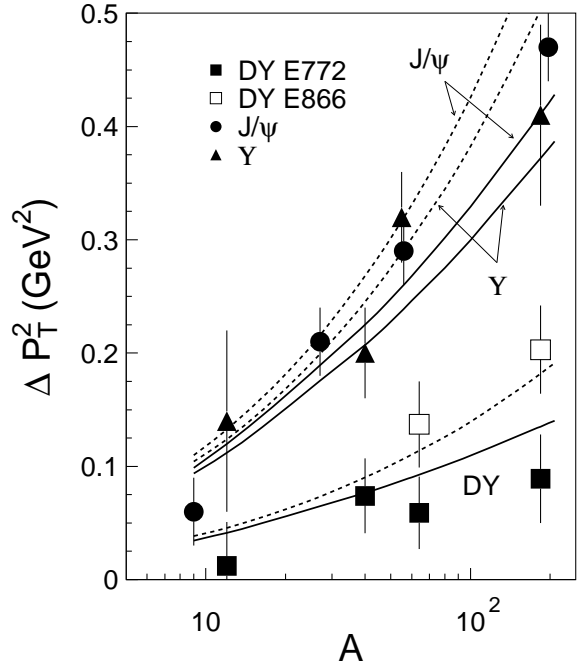}
 \caption{\label{broad} Broadening in 
Drell-Yan reaction on different nuclei as measured in the E772  (closed squares) \cite{e772}
and E866 (open squares) \cite{e866-pt} experiments respectively.  Broadening for $J/\Psi$ and $\Upsilon$ \cite{e772,mmp} is shown by circles and triangles respectively.
The dashed and solid curves correspond to the predictions without and with the corrections for gluon shadowing respectively.}
 \end{figure}

 Radiation of prompt photons and dileptons should be a sensitive probe for 
 broadening of projectile quarks, as was stressed in Sect.~\ref{photons}. While broadening of direct photons is difficult to measure, since the small $p_T$ region is overwhelmed by radiative hadronic decays,
data for heavy dileptons are available. Figure~\ref{broad} shows the results of the E772 and E866 fix target experiments at Fermilab for broadening in Drell-Yan reaction at $800\GeV$.

To calculate broadening of heavy dileptons we used the results of Sect.~\ref{photons}. The mean value of the fractional light-cone momentum of detected dileptons in the E772 experiment was $\la x_1\ra=0.26$ \cite{peng}. Accordingly, the coherence length is sufficiently short to rely on the equation (\ref{277}).

Broadening for a Drell-Yan pair is factor $z^2$ smaller than that for the projectile quark which radiated the heavy dilepton, where $z$ is the fraction of the quark momentum carried by the dilepton. We use here the mean value $\la z\ra=0.9$, as was evaluated in \cite{krtj}. We calculated the factor $C(E)$ at the energy $E=\la x_1\ra s/2m_N\la z\ra$.

The results for Drell-Yan reaction are shown in Fig.~\ref{broad} by the bottom curve.
Notice that our result for Drell-Yan reaction is close to broadening calculated in \cite{jkt},
which was about twice as big as the broadening previously observed in the E772 experiment (closed squares in Fig.~\ref{broad}). However, data from the E866 experiment (open squares in Fig.~\ref{broad}) published later confirmed the predictions made in \cite{jkt}.

\subsection{Heavy quarkonium production}\label{psi}

We assume that the energy of the projectile gluon is equal to the energy of the heavy quarkonium it produces, i.e. $E_g=x_1\,s/2m_N$, where $x_1$ is the fractional light-cone momentum of the quarkonium. Indeed, the fraction of the gluon momentum carried by the quarkonium is model dependent, and is either  equal \cite{ryskin} or very close to one \cite{color-singlet}.

Since broadening was measured for a large sample of events, we rely on the mean values of 
$x_1$, which were in the E772 experiment, $\la x_1\ra=0.29$ and $0.23$ for $J/\Psi$ and $\Upsilon$ respectively  \cite{peng}.  At such energy of the gluon, $E=\la x_1\ra s/2m_N$, gluon radiation occurs
coherently, i.e. the lifetime of the projectile gluon is sufficienly long to propagate through the whole nucleus. However the coherence time for heavy quarkonium production is short, i.e. the projectile gluon converts into the quarkonium almost instantaneously. Therefore, we should rely on Eq.~(\ref{270}), but using about half of the nuclear thickness, like in the Drell-Yan case.
However, heavy quarkonia have a nonzero absorption cross section. This makes the path length available for broadening a bit longer. Broadening is proportional to the amount of nuclear matter passed by the projectile gluon prior production of a heavy quarkonium at the point with longitudinal coordinate $z$, which equals to,
\beq
T_z(b)=\int\limits_{-\infty}^z dz'\,\rho_A(b,z)
\label{300}
\eeq
The mean value of $T_z(b)$ reads,
\beqn
\left\la T_z(b)\right\ra&\equiv&
\frac{
\int_{-\infty}^{\infty}dz\,
\rho_A(b,z)\,T_z(b)\,
e^{-\sigma_{abs}[T_A(b)-T_z(b)]}}
{\int_{-\infty}^{\infty}dz\,
\rho_A(b,z)\,
e^{-\sigma_{abs}[T_A(b)-T_z(b)]}}
\nonumber\\ &=&
\frac{T_A(b)}{1-e^{-\sigma_{abs}T_A(b)}} -
\frac{1}{\sigma_{abs}},
\label{320}
\eeqn
where $\sigma_{abs}$ is the inelastic cross section of the produced heavy quarkonium on a nucleon. This cross section is small, so the magnitude of $\la T_z(b)\ra$ is not much different from $T_A(b)/2$. Therefore we can simplify the averaging over impact parameter, by replacing $T_A(b)\Rightarrow \la T_A\ra$. We also neglect the color transparency effects \cite{kz91,hikt2}
using  the physical values of $\sigma_{abs}$, which we fix at $5\mb$ for $J/\Psi$ and a factor 
$(m_c/m_b)$ less for $\Upsilon$ \cite{hikt1}.

The results of calculations are compared in Fig.~\ref{broad} with data for broadening of $J/\Psi$ and $\Upsilon$ production measure in the E772 experiment \cite{e772,mmp}. The two upper curves differ from each other, since $J/\Psi$ and $\Upsilon$ have different absorption cross sections and slightly different values of $\la x_1\ra$ \cite{peng}. Agreement with data again is rather good.

Notice that in order to calculate nuclear broadening for heavy quarkonium production one does not need to know its mechanism  provided that the coherence length of quarkonium production is short, 
\beq
l_c=\frac{s x_1}{m_N M_{\bar QQ}^2}\ll R_A.
\label{330}
\eeq
 The data for $\Upsilon$ production from the E772 experiment depicted in Fig.~\ref{broad} satisfy well this condition. However, the data for $J/\Psi$ production are somewhat out of this kinematic domain.
 Therefore we should rely mostly on the comparison with the data for $\Upsilon$ production. Nevertheless, the data for both quarkonia look very similar in Fig.~\ref{broad}, suggesting a weak dependence of broadening on $l_c$.  

Data at higher energies are available at RHIC. Broadening for $J/\Psi$ production in deuteron-gold collisions at $\sqrt{s}=200\GeV^2$ \cite{phenix-psi} was measured at medium, backward and forward rapidities, demonstrating no clear dependence on rapidity.  The measured magnitude of broadening agrees with the E772 data within rather large errors.
These observations confirm the weak $l_c$ dependence of broadening in heavy quarkonium production.

\subsection{Broadening in SIDIS}

Another source of experimental information on quark broadening is hadron production in semi-inclusive deep-inelastic scattering (SIDIS). Broadening of produced hadrons was measured recently by the HERMES collaboration at HERA \cite{hermes} and by the CLAS collaboration at Jefferson Lab \cite{jlab}. This reaction has some advantages compared to Drell-Yan process, since it has more certain kinematics. Indeed, at large Bjorken $x\gsim0.1$ the whole energy of the virtual photon is transferred to the quark, which hadronizes to a leading hadron with a measurable fraction $z_h$ of its light-cone momentum. 

At these values of $x$ the SIDIS processes on different nucleons do not interfere and a part of nuclear broadening comes from Fermi-motion of the bound nucleon. Its contribution to hadron broadening is easy to evaluate,
\beq
\left(\Delta p_T^2\right)_F={2\over3}\,x^2\,z_h^2\,\la p_F^2\ra,
\label{335}
\eeq
Where $\la p_F^2\ra\sim 0.04\GeV^2$ is the mean Fermi momentum squared. Thus, this correction is well under control and in most cases is very small and can be neglected.

Unfortunately, broadening in SIDIS suffers of considerable theoretical uncertainties and model dependence. Indeed, the quark knocked out of a bound nucleon propagates through the nuclear medium and experiences broadening only until its color is neutralized and a colorless pre-hadron is produced \cite{knp,knph,trieste}. For leading hadron production the pre-hadron rescatterings can occur only with small elastic cross section and can be disregarded. The production length $l_p$ of the pre-hadron
can be only calculated within models and is less known than broadening. Model calculations \cite{pir} are in a reasonable agreement with HERMES data, but this should be treated as a test of our knowledge of $l_p$.

To get rid of this uncertainty one can go to higher energies, since $l_p$ rises linearly with anergy (at fixed $Q^2$) and eventually one may think that all pre-hadrons are produced outside the nucleus.
However, another problem immediately emerges:   at high energies one gets into the region of small $x$ dominated by dijet production. Then the photon energy is shared by the produced quark and antiquark jets, and one does not know from which jet originated the detected hadron. This means that the fractional hadron momentum $z_h$ is not known any more, so broadening of the hadron cannot be easily translated to broadening of the quark.

Nevertheless, one can relate the broadenings of quark and produced hadron on a rather firm theoretical basis. This relation has the form,
\begin{widetext}
\beqn
\Delta \left(p_T^2\right)_h &=&
\frac{z_h^2\,\left(\Delta p_T^2\right)_q}{\int d^2r_T \int_0^1 d\alpha |\Psi_{\gamma^*}(r_T,\alpha)|^2
\,\sigma_{\bar qq}(r_T,{ x})}
\times \int d^2r_T
\Biggl\{\int\limits^1_{ z_h}\frac{d\alpha}{\alpha^2}\,
\left|\Psi_{\gamma^*}(r_T,\alpha,{ Q^2})\right|^2
\sigma_{\bar qq}(r_T,{ x})D_{h/q}\left(\frac{{ z_h}}{\alpha},{ Q^2}\right)
\nonumber\\ &+&
\int\limits_0^{1-{ z_h}} \frac{d\alpha}{(1-\alpha)^2} \,
\left|\Psi_{\gamma^*}(r_T,\alpha,{ Q^2})\right|^2
\sigma_{\bar qq}(r_T,{ x})D_{h/\bar q}\left(\frac{ z_h}{1-\alpha},{ Q^2}\right)\Biggr\}.
\label{400a}
\eeqn
\end{widetext}
Here the photon of virtuality $Q^2$ is assumed to convert into a $\bar qq$ pair with fractional momenta $\alpha$ and $1-\alpha$, which distribution amplitude $\Psi_{\gamma^*}(r_T,\alpha,{ Q^2})$.
The latter is well known from QED \cite{bjorken}. The phenomenological dipole cross section
$\sigma_{\bar qq}(r_T,{ x})$ and the quark fragmentation function $D_{h/q}(z_h,Q^2$ are well fitted to HERA data for the proton structure function and to data on jet fragmentation in $e^+e^-$ annihilation.

Thus, data on broadening in SIDIS taken at future electron-ion colliders  should bring forth precious information on quark broadening in nuclei.

\subsection{Cronin effect}

In hadron-nucleus collisions projectile partons also experience broadening propagating through the nuclear target. This leads to the so called Cronin effect, nuclear enhancement of particle production at medium high momentum transfers. The dipole formalism described  in Sect.~\ref{broadening} describes well the data on pion production in $pA$ collisions in a parameter free way \cite{cronin}. Moreover,
this formalism correctly predicted the weak Cronin enhancement, $\sim 10\%$, observed later at RHIC (see in \cite{spots}).
Even a smaller effect, partially compensated by gluon shadowing, is expected at LHC \cite{cronin,last-call}.

Cronin effect was also observed in Drell-Yan reaction \cite{e772,e866,e866-pt} and is also well explained quantitavely by the dipole formalism \cite{mikkel}.

\section{Models for saturation confront data for broadening}

The saturation scale in the eikonal approximation was predicted in \cite{kn,kl,kln,hirano} to be,
\beq
Q_{gA}^2(x,b)= 
\frac{3\pi^2}{2}\,\alpha_s(Q_{gA}^2)\,xG_N(x,Q_{gA}^2)\,\rho_{part}^A(b),
\label{440}
\eeq
where  $x=Q_{gA}^2/2m_NE$ and $\rho_{part}^A(b)$ is the number of participants. For $pA$ collisions  $\rho_{part}^A(b)=T_A(b)$ and this equation corresponds to our relation Eq.~(\ref{130}), written at $r_T=1/Q_{gA}$, and enhanced for gluons  by the Casimir factor $9/4$.

The saturation scale Eq.~(\ref{440}) was calculated in \cite{hirano} for central gold-gold collisions using the gluon density $xG(x,Q^2)=0.7\,\ln(1+Q^2/\Lambda_{QCD}^2)\, x^{-0.2}(1-x)^4$.
The result was applied to obtain hadron multiplicities in heavy ion collisions, and good agreement with data from RHIC was found.

\begin{figure}[htb]
\includegraphics[width=7cm]{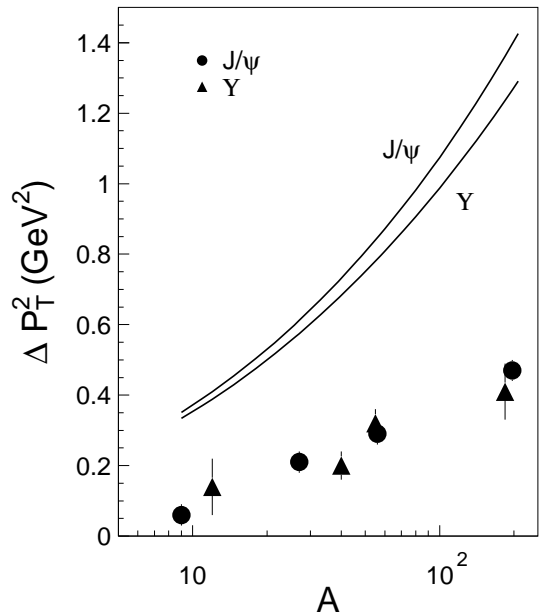}
 \caption{\label{hirano} Comparison of the saturated scale Eq.~(\ref{440}) calculated in \cite{hirano} with data \cite{e772,mmp} on broadening of $J/\Psi$ and $\Upsilon$ production in $pA$ collisions at $800\GeV$. }
 \end{figure}

Apparently, such a comparison with data cannot be considered as a rigorous test of the model Eq.~(\ref{440}), since the dynamics of particle production in nuclear collisions is very complicated (unknown) and involves many additional assumptions.  Instead we perform here a direct test of Eq.~(\ref{440}), comparing it with broadening of heavy quarkonia, like we did in Sect.~\ref{psi}. Fig.~\ref{hirano}
demonstrates comparison of the saturation scale calculated in \cite{hirano} with broadening of $J/\Psi$ and $\Upsilon$ production measured in $pA$ collisions at $E_p=800\GeV$ in the experiment E772 \cite{e772,mmp}.

We see that the model grossly, more than 3 times,  overestimates the data. The source of disagreement is in the choice of the effective scale $Q^2=Q^2_{gA}$ in the right-hand side of (\ref{440}), which might look natural, but is ill justified. 

At this point we should notice that the large mean transverse momentum squared $\la\p_T^2\ra=Q_{gA}^2$ has its origin in many soft interactions with a tiny mean momentum transfer in each of them \cite{jkt}. Therefore, one should use a soft scale, rather than $Q_{gA}$, in the eikonal formula Eq.~(\ref{440}). This is of course a risky procedure, and is better to rely on phenomenology which effectively incorporates unknown soft dynamics. 

The sharing of the momentum transfer between multiple interactions changes when one overcomes the saturation scale at $p_T^2\gg Q_{gA}^2$.  In this case, due to the power $p_T$-dependence of a single interaction cross section, the main contribution comes from a single scattering with large momentum transfer \cite{levin-rysk}. In this case the scale is indeed controlled by $p_T$, but not in Eq.~(\ref{440}).

One can also treat Eq.~(\ref{440}) as an implicit equation for $Q_{gA}$ \cite{kn}. It turns out, however,
that this equation with a realistic gluon density, e.g.  the recent result of the phenomenological analysis of data MSTW2008 \cite{mstw},
 does not have a solution. As a result of an iteration procedure the value of $Q_{gA}^2$ is drifting to very small values, out of the range of applicability of the MSTW2008 code. This happens because contemporary 
 analyses show a very small gluon density at low scale dictated by the latest measurements of DIS at small virtualities. This is another manifestation of small gluonic spots in the proton \cite{spots}: gluons are invisible for measurements with low resolution $Q^2$.

\section{Summary}

\begin{itemize}

\item
The saturation scale for partons in nuclei is directly related to the magnitude of transverse momentum broadening of a parton propagating through the nucleus.

\item
We employed the light-cone dipole approach to predict the magnitude of broadening. We found that broadening strongly depends on the coherence length of the process and is different for simple propagation of a parton and for a parton splitting processes. 

\item
The same LP effect which causes saturation, leads to a reduction of gluon density in nuclei, what in turn reduces broadening. We derived Eq.~(\ref{279e}), which quantifies this effect.

\item
Using these results we predict and compare with data broadening in Drell-Yan reaction, as well as in production of heavy quarkonia. The results of our parameter-free calculations are in good agreement with data, plotted in Fig.~\ref{broad}.  

\item
Although broadening of hadrons produced in SIDIS suffer from a considerable theoretical uncertainty in the description of the space-time development of in-medium hadronization, we expect that such measurements on future electron-ion colliders will become a precious source of information about quark broadening at high energies.

\item
We found that the saturation momentum which describes well the multiplicity of hadrons produced in heavy ion collisions is far too high compared with the transverse momentum broadening  measured in $pA$ collisions.

\end{itemize}

 \begin{acknowledgments}

We are thankful to Will Brooks for his interest and useful discussion, and to Jen-Chieh Peng for providing information on important details of measurements performed in the E772 experiment.
This work was supported in part by Fondecyt (Chile)
grants 1090236 and 1090291, and by DFG (Germany) grant PI182/3-1.

\end{acknowledgments}

\end{document}